\documentclass[11pt]{article}
\usepackage[utf8x]{inputenc}
\usepackage{amsmath}
\usepackage{graphicx}
\usepackage{geometry}
\usepackage{float}
\usepackage[hang, small]{caption}
\newcommand{\bfuwe}[2]{#1}

\begin{document}
\title{\textbf{From a thin film model for passive suspensions towards the description of osmotic biofilm spreading} }
\author{Sarah Trinschek$^{1,2}$, Karin John$^{2}$ and  Uwe Thiele$^{1,3\, *}$  \\
\\
\small{$^{1}$ Institut f\"ur Theoretische Physik, Westf\"alische Wilhelms-Universit\"at M\"unster,}  \\ \small{Wilhelm Klemm Strasse 9, 48149 M\"unster, Germany} \\
\small{$^{2}$ Laboratoire Interdisciplinaire de Physique (LIPhy), CNRS / Universit\'e Grenoble-Alpes,}   \\ \small{140 Rue de la Physique, 38402 Grenoble, France}\\
\small{$^{3}$ Center of Nonlinear Science (CeNoS), Westf\"alische Wilhelms-Universit\"at M\"unster,}  \\ \small{Corrensstr. 2, 48149 M\"unster,  Germany}\\
\\
\small{ \textbf{$^*$ Correspondence:} u.thiele@uni-muenster.de }
}
\date{}
\maketitle

\begin{abstract}
Biofilms are ubiquitous macro-colonies of bacteria that develop at \bfuwe{various interfaces (solid-liquid, solid-gas or liquid-gas).} The formation of biofilms starts with the attachment of individual bacteria \bfuwe{to an interface}, where they proliferate and produce a slimy polymeric matrix - two processes that result in colony growth and spreading.
Recent experiments on the growth of biofilms on agar substrates under air have shown that for certain bacterial strains, the production of the extracellular matrix and the resulting osmotic influx of nutrient-rich water from the agar into the biofilm are more crucial for the spreading behaviour of a biofilm than the motility of individual bacteria.
We present a model which describes the biofilm evolution and the advancing biofilm edge for this spreading mechanism. \bfuwe{The model is based on a gradient dynamics formulation for thin films of biologically passive liquid mixtures and suspensions, supplemented by bioactive processes which play a decisive role in the osmotic spreading of biofilms. It explicitly includes the wetting properties of the biofilm on the agar substrate via a disjoining pressure and can therefore give insight into the interplay between passive surface forces and bioactive growth processes.}

\end{abstract}
\maketitle

\section{Introduction}

The formation of biofilms \bfuwe{at solid substrates} is initiated by the attachment of individual bacteria to a surface \cite{Donlan2002EID}. These attached bacteria then proliferate and at the same time produce a slimy polymeric matrix. This extracellular matrix protects the bacteria and also gives the biofilm its mechanical properties, which govern its morphogenesis and allow it to resist mechanical stresses \cite{FW2010NRM}. As proliferation and matrix production proceed, the colony grows and spreads laterally along the substrate. \bfuwe{This process may occur in an ambient liquid or under a gas atmosphere. }

\bfuwe{The widespread occurrence of biofilms and their either detrimental or beneficial effect imply that it is highly important to understand the physico-chemical and biological principles of biofilm formation and spreading at different types of interfaces. Hence the past decades have seen the development of a variety of theoretical biofilm models.  On the one hand, there is a vast literature on the evolution of biofilms on flat solid substrates in contact with a bulk liquid, i.e., at solid-liquid interfaces (for reviews see for example \cite{PL2003, WZ2010SSC, Horn2014}). Independent of the type of model used (continuous, discrete or hybrid), the following processes are typically considered: cell attachement/detachment to/from the surface, cell division and cell death, matrix production, nutrients and oxygen uptake by the cell etc. Transport is modeled via diffusion and convection through the liquid phase. Relevant questions, which have been treated are e.g. the effect of cell transport 
mechanisms \cite{
PVH+1998bab, EPV2001CaMMiM}, nutrient concentration \cite{WC1997fme, Hermanowicz2001mb}, biofilm matrix properties \cite{CK2004MMB, ZCW2008sjoam, ZCW2008ccp, WCM+2011}, or cell-to-cell signalling \cite{WK2012JEM} on the biofilm morphology, the interplay between bulk fluid flow and biofilm morphology \cite{DFF+1996JoCP, PVH2000BaB}, or the evolution of multispecies biofilms \cite{Wanner2000,PKL2004Aaem, AK2007BoMB}. }

\bfuwe{On the other hand, recently, also biofilms at solid-gas interfaces have attracted some attention as in the laboratory and in nature, biofilms also grow on moist solid substrates in contact with a gas phase. In this case, the spreading of a biofilm colony involves the motion of the three-phase contact line between the viscous biofilm, the solid substrate and the gas phase. In the contact line region wetting phenomena are likely to play a major role in the biofilm dynamics. Indeed, it was shown experimentally, that in many cases biofilm spreading is not driven by the active mobility of individual bacteria but rather by growth processes and the physico-chemical properties of the biofilm components and the resulting fluxes \cite{ DML2014IF,SAW+2012PNASUSA, DTH2014PotRSoLBBS, DFM+2015, FPB+2012SM, ARK+2009PNASU, Brenner2014fluid}. One example is a spreading mechanism studied for \textit{Bacillus subtilis} biofilms which relies on the auto-production of surfactin, which acts as 
surfactant.  Surfactants can promote 
rapid surface colonisation by altering the interfacial properties \cite{DFM+2015, Leclere2006}. Furthermore, a local gradient in surfactant concentration results in a surface tension gradient which drives so-called Marangoni flows that promote spreading \cite{FPB+2012SM, ARK+2009PNASU}. Another example for a biofilm spreading mechanism that corresponds to a physico-chemical mechanism is the osmotic spreading. }
\bfuwe{Osmosis describes the transport of solvent across an interface that the solute is not able to pass. The
most emblematic example is that of the transport of water through a semi-permeable membrane
from a reservoir with low solute concentration into one with high solute concentration \cite{Lach2007ajp}.
For \textit{Bacillus subtilis} and \textit{Sinorhizobium meliloti}, the spreading behaviour on moist agar substrates is determined by the ability of the bacteria to produce the exopolysaccharide component of the polymeric matrix \cite{SAW+2012PNASUSA, DTH2014PotRSoLBBS}. The spreading mechanism results from gradients of osmotic pressure that are generated as bacteria consume water and nutrient to produce further biomass. The resulting osmotic imbalance between biofilm and agar substrate causes the biofilm to swell and spread through an uptake of water from the moist agar substrate. It was found experimentally, that nutrient depletion within a colony that grows on a nutrient-rich agar substrate in a gas atmosphere triggers an increase in matrix production that allows for a subsequent radial spreading which provides the biofilm with fresh nutrients \cite{ZSS+2014NJoP}. Recently, experimental tracking of the distribution of the main phenotypes of a \textit{Bacillus subtilis} strain with fluorescence 
imaging techniques has shown that the matrix producing cells are predominantly located in the outer rim of the biofilm which also suggests that they play a crucial role in the spreading of a biofilm front \cite{Wang2016AMB}. }

\bfuwe{ As outlined above, many aspects of the complex biofilm morphology have been modeled extensively with very detailed multidimensional approaches.  With our present approach we are specifically aiming at investigating the role of surface forces, i.e.\ capillarity and wettability, on the osmotic spreading of biofilms growing at agar-gas interfaces. Thereby we neglect many aspect of biofilm complexity (for example specifically modelling the nutrient and oxygen dynamics). Instead we start out from the thermodynamic consistent theoretical framework developed for the evolution of films of passive suspensions \cite{TTL2013prl,Thiele2011ES,XTQ2015JPCM}, which we supplement by bioactive growth terms representing proliferation/death and matrix production. In the first presentation of this approach we focus on a simple case, where the biomass is only transported by passive mechanisms, as opposed to active transport via bacterial motility.  Because the osmotic spreading mechanism of biofilm growth results 
from physico-chemical processes occurring at the interface between the agar substrate and the biofilm, we include wetting effects into the model in terms of a Derjaguin (or disjoining) pressure that is related to a wetting energy. 
This approach allows us to consistently describe the advancing biofilm edge and to study how biofilm growth results from the interplay between passive surface forces (i.e., wetting energy, surface tension), osmotic fluxes between agar and biofilm, and active growth processes.}

\bfuwe{The following section~\ref{sec:mm} presents our model and its numerical treatment. First, section~\ref{sec:mod-pass} introduces the model for the passive case of a free-surface film of a suspension on a solid substrate in a gradient dynamics form. It describes the spreading film in terms of coupled evolution equations for continuous variables, namely, the effective heights of solute and solvent. We emphasise that the gradient dynamics form (i) allows for an easy and fully consistent extension towards more complex materials incorporating, e.g., solvent-solute and solute-solute interactions \cite{Thiele2011ES,XTQ2015JPCM,ThAP2012pf}, (ii) directly reflects the principles of non-equilibrium thermodynamics \cite{GrootMazur1984}, and (iii) and is fully equivalent to the hydrodynamic form obtained by long-wave approximation that consist of a thin film equation coupled to an advection-diffusion equation \cite{TTL2013prl,XTQ2015JPCM}. The subsequent section~\ref{sec:mod-act} introduces the bioactive growth 
term into the model and the term responsible for the osmotic influx and evaporation of the solvent. From there on, solvent and solute represent nutrient-rich water and biomass (bacteria and the extracellular polymeric matrix), respectively. } \bfuwe{Our numerical treatment of the model is briefly discussed in section~\ref{sec:num}.}

\bfuwe{Section~\ref{sec:res} presents selected results obtained with our simplified model. First,  we show that the model qualitatively reproduces the experimentally observed transition of growing biofilm droplets from initial vertical swelling to subsequent horizontal spreading \cite{SAW+2012PNASUSA}. At later times, the biofilm spreads with a front of constant shape and velocity. Next, we study how the front velocity depends on relevant model parameters, e.g., wettability of the substrate, biomass growth rate constant, transport coefficients, etc. and compare the observed tendencies with available data in the literature. We also present a first simulation of the osmotic growth of a biofilm on a fully two-dimensional substrate. The final section~\ref{sec:conc} presents our conclusion. }

\section{Materials and Method}
\label{sec:mm}
To study the osmotically driven spreading mechanism of biofilms, we develop a basic mathe\-ma\-ti\-cal model that predicts the dynamics of a biofilm on a flat moist substrate. 
From the viewpoint of soft condensed matter
physics, biofilms may be considered as \bfuwe{thin films of suspensions of bioactive colloids in
  viscoelastic liquid \cite{WAS+2011MB}. In consequence, thin-film descriptions were recently} introduced for biofilms to study early stage biofilm-growth and quorum sensing \cite{WK2012JEM}, osmotically driven spreading \cite{SAW+2012PNASUSA} and the effect of surfactant production for the spreading of a bacterial colony that moves up a non-nutritive boundary \cite{ARK+2009PNASU}.

\begin{figure}[h]
\begin{center}
\includegraphics[width=0.45\textwidth]{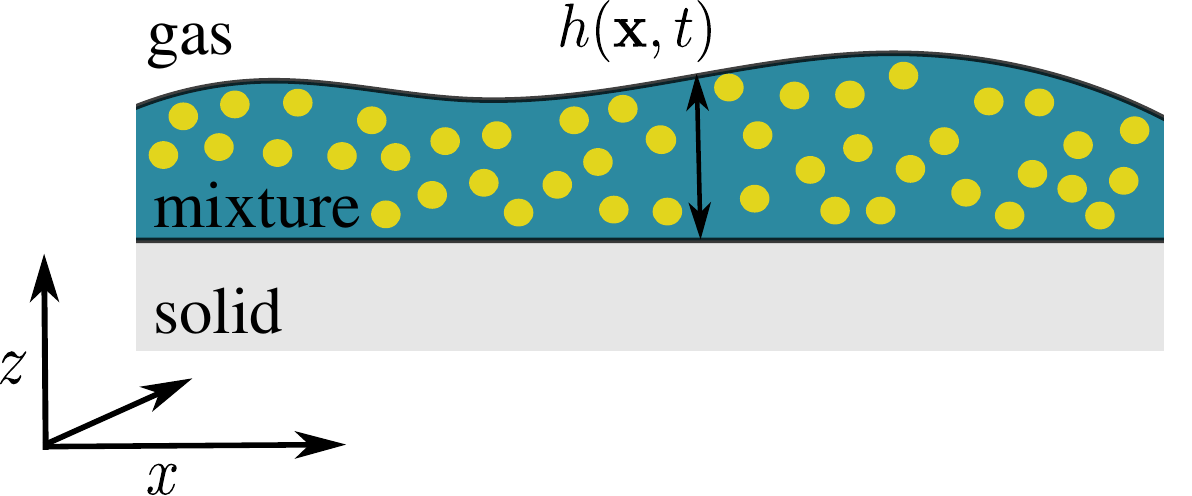}  
\caption{Sketch of a thin film of binary mixture on a flat solid substrate. The total film height $h(\textbf{x},t)=\psi_1(\textbf{x},t)+\psi_2(\textbf{x},t)$ is given by the sum of the effective layer thicknesses of solvent $\psi_1(\textbf{x},t)$ (nutrient-rich water) and solute $\psi_2(\textbf{x},t)$ (bacteria and extracellular matrix).}
\label{Fig1}
\end{center}
\end{figure}
 
We assume a biofilm of height $h$ consists of a mixture of nutrient-rich water with the height-averaged concentration $\phi_1 = \frac{\psi_1}{h}$ and of biomass (bacteria and the extracellular polymeric matrix) with the height-averaged concentration $\phi_2 = \frac{\psi_2}{h}$ (see sketch in Fig.~\ref{Fig1}). 
\bfuwe{We base our model on the fields $\psi_1$ and $\psi_2$ that denote the effective layer thickness of water and biomass, respectively. The total film height
can always be obtained via $h=\psi_1+\psi_2$. }
\bfuwe{In section~\ref{sec:mod-pass}, the passive aspect of the biofilm dynamics is modeled using the thermodynamically consistent gradient dynamics formalism developed for the description of films of binary mixtures and suspensions \cite{TTL2013prl,Thiele2011ES,XTQ2015JPCM}. To facilitate comparison to the literature, the hydrodynamic long-wave formulation of the model is also given. 
In section~\ref{sec:mod-act}, the gradient dynamics is supplemented by bioactive terms which account for processes that play a crucial role for the osmotic spreading of biofilm droplets. Hereby we assume that in the early stages of biofilm development, the hydrodynamic coupling of the individual bacteria is negligible. }
 
\subsection{Modeling thin films of passive mixtures and suspensions}
\label{sec:mod-pass}
Under the assumption that variations in film height are small as compared to all relevant horizontal length scales, the behaviour of free surface layers and shallow drops of simple liquids can be described in terms of an evolution equation for the film thickness profile $h(\textbf{x},t)$ in the gradient dynamics form $\partial_t h = \nabla \cdot [  Q(h) \nabla \frac{\delta F[h]}{\delta h}] $ with a mobility function $Q(h)$ and a free energy functional $F[h]$ containing interface and wetting energies \cite{Mitl1993jcis, Thiele2010JPCM}. \bfuwe{With the mobility $Q=h^3$, the gradient dynamics formulation is fully equivalent to the standard thin film equation obtained via long-wave approximation from the governing equations of hydrodynamics \cite{ODB1997RMP,Thiele2007}. }

Considering now a thin free-surface film of a binary suspension or solution on a flat solid substrate, one can extend this approach and derive two coupled evolution equations for the effective layer thicknesses $\psi_i$ of solvent and solute \cite{XTQ2015JPCM, WTG+2015apa}:
 \begin{align}
 & \partial_t \psi_1 = \nabla \cdot \left[Q_{11} \nabla \frac{\delta F}{\delta \psi_1} +Q_{12} \nabla \frac{\delta F}{\delta \psi_2}        \right]  \notag \\
 & \partial_t \psi_2 = \nabla \cdot  \left[Q_{21} \nabla \frac{\delta F}{\delta \psi_1} +Q_{22} \nabla \frac{\delta F}{\delta \psi_2}        \right] \, .
  \label{mixtureeqn}
\end{align}
The underlying free energy functional
 \begin{equation}\label{eq:en}
  F[\psi_1, \psi_2] = \int \, [ \,   \underbrace{ f(\psi_1, \psi_2)}_\text{wetting} + \underbrace{ (\psi_1+ \psi_2) g(\psi_1, \psi_2)}_\text{bulk mixing}+  \underbrace{\tfrac{\gamma}{2} (\nabla (\psi_1+ \psi_2))^2}_\text{surface} \,   ] \,   \mathrm{dx} 
 \end{equation}
consists of the wetting energy  $f(\psi_1, \psi_2)$, a film bulk contribution $ g(\psi_1, \psi_2)$ that accounts for the entropic terms and the interactions of solute and solvent, and a surface term that depends on the interface shape and the liquid-gas interface tension $\gamma$. In general, these contributions can be chosen to include composition-dependent wetting properties and complex solvent-solute interactions. 
Here, we restrict ourselves to relatively simple choices: The wetting energy depends only on the overall film height $h = \psi_1 + \psi_2$ \bfuwe{and describes a biofilm that partially wets the substrate, i.e., in the contact region the film-air interface forms a small finite angle with the substrate \cite{deGennes1985}. A simple 
such energy is
\begin{equation}
  f(\psi_1, \psi_2) = \tilde  f(\psi_1+\psi_2) = \left( - \frac{A} {2(\psi_1+ \psi_2)^2} + \frac{B}{5 (\psi_1+ \psi_2)^5} \right) \quad \text{with} \, A,B \text{=const} \label{djp}
\end{equation}
that combines long-range van-der-Waals interactions characterised by the Hamaker constant $A$ and stabilising short range interactions characterised by the constant $B$. The usual Derjaguin (or disjoining) pressure is obtained as $-d\tilde f /d h$. The constants may be obtained from the contact angle and the thickness of the wetting layer on the moist agar that coexists with the film  \cite{Israelachvili1986BdBfupC,GennesBrochard-WyartQuere2004}. Eq.~(\ref{djp}) is a common choice that together with the other contributions in $F[\psi_1, \psi_2]$  well describes spreading or dewetting behaviour of passive droplets \cite{Thiele2007, Bonn2009}. Other choices of $\tilde  f$ are possible, but do normally not change the qualitative behaviour. }

The bulk contribution $g$ in Eq.~(\ref{eq:en}) represents the free energy of mixing of solute and solvent. In the case of purely entropic contributions we have
 \begin{equation}\label{eq:gg}
 g(\psi_1, \psi_2)= \frac{k_\text{B}T}{a^3}  \left[ \frac{\psi_1}{(\psi_1+ \psi_2)}  \ln \left(\frac{\psi_1}{(\psi_1+ \psi_2)}\right)    + \frac{\psi_2}{(\psi_1+ \psi_2)}  \ln \left(\frac{\psi_2}{(\psi_1+ \psi_2)}\right)     \right]  \, ,
 \end{equation}
 where $k_\text{B}T$ denotes the thermal energy \bfuwe{and $a$ is a typical length scale related to the solvent particle (e.g., related to the size of the bacteria  and the constituents of the extracellular polymeric matrix). The incorporation of further interactions between solute and solvent into $g$ is straightforward \cite{XTQ2015JPCM} is, however, not considered here. }
The symmetric and positive definite mobility matrix
\begin{equation}
 \mathbf{Q}(\psi_1, \psi_2) = \begin{pmatrix} Q_{11} & Q_{12} \\ Q_{21} & Q_{22} \end{pmatrix} = \underbrace{D \frac{\psi_2 \psi_1}{\psi_1+ \psi_2  }  \begin{pmatrix} 1 & -1 \\ -1 & 1 \end{pmatrix}}_\text{diffusion} + \underbrace{ \frac{\psi_1+ \psi_2}{3  \hat{\eta}(\psi_1,\psi_2)}   \begin{pmatrix} \psi_1^2 & \psi_2 \psi_1 \\ \psi_2 \psi_1 & \psi_2^2 \end{pmatrix} }_\text{convection} \label{mix:mob}
\end{equation}
contains contributions from diffusive and convective transport and can be obtained via a long-wave approximation of the full hydrodynamic model as discussed in \cite{TTL2013prl,Thiele2011ES,XTQ2015JPCM,ThAP2012pf,WTG+2015apa,NaTh2010n}. It depends on the viscosity of the solution $\eta(\psi_1,\psi_2) = \eta_0 \hat{\eta}(\psi_1,\psi_2)$, which is written in terms of a reference viscosity $\eta_0$ (i.e., the viscosity of the nutrient-rich water) and a dimensionless scaling function $\hat{\eta}(\psi_1,\psi_2)$ that captures the composition-dependence of the viscosity. The viscosity of the biomass $\eta_b$ is about 10-1000 times higher than the viscosity of the bulk liquid $\eta_0$ \cite{HCS2004NRM, Sutherland2001m,LDB+2009bj}. We use the linear ansatz
 \begin{equation}
 \hat{\eta}(\psi_1,\psi_2) =\frac{\psi_1}{h} + \frac{\eta_b}{\eta_0} \frac{\psi_2}{h} \label{vis}
 \end{equation}
which directly enters the convective part of the mobility matrix $\mathbf{Q}(\psi_1,\psi_2)$ where it primarily affects the time-scale of biofilm spreading dynamics.
\bfuwe{The diffusivity is $D= \frac{a^2}{6 \pi \eta}$ and is consistent with the Einstein relation between diffusion constant $D_\text{diff}= D \frac{k_\text{B} T}{a^3} = \frac{k_\text{B} T}{6 \pi a \eta} $ and viscosity. }

Eqs.\ (\ref{mixtureeqn}-\ref{vis}) describe the evolution of a passive mixture of solute and solvent under the influence of entropic and surface forces. For a partially wetting \bfuwe{suspension-substrate-air combination with the wetting energy in Eq.~(\ref{djp}) it results in steady mixture droplets sitting on a wetting layer of height $h_p^3=\tfrac{B}{A}$. } The dynamics of the (macroscopic) contact line between the substrate, the gas phase and the biofilm is naturally encoded in the evolution equations and no additional assumptions on the relation between forces and the contact line velocity have to be made.

\bfuwe{Before we discuss the employed scaling, we present Eqs.~(\ref{mixtureeqn}-\ref{vis}) in the usual hydrodynamic form of a thin-film model. A variable transformation is used to go from the presented description in terms of the effective solvent and solute heights $(\psi_1,\psi_2)$ to an alternative form of the same model written in terms of the film height $h=\psi_1+\psi_2$ and height-averaged biomass concentration $\phi_2=\psi_2/(\psi_1+\psi_2)$. Introducing then the specific forms of $\tilde f$ and $g$ [Eqs.~(\ref{djp}) and (\ref{eq:gg}), respectively] into the energy (\ref{eq:en}) and then together with (\ref{vis}) into Eqs.~(\ref{mixtureeqn}) results in the classical form of a thin film equation coupled to an advection-diffusion equation 
 \begin{align}
 & \partial_t h = \nabla \cdot \mathbf{j}_\mathrm{conv} \quad\mathrm{and}\quad
\partial_t (h\phi_2) = \nabla \cdot  (\phi_2\mathbf{j}_\mathrm{conv}+\mathbf{j}_\mathrm{diff} ) \notag \\
 & \mathrm{with}\qquad \mathbf{j}_\mathrm{conv}=-\frac{h^3}{3\eta}\nabla\left(\gamma\Delta h - \frac{d\tilde f}{dh}\right) \quad\mathrm{and}\quad
\mathbf{j}_\mathrm{diff}  = D_\mathrm{diff} h \nabla\phi_2\, .
  \label{mixturehydro}
\end{align}
known from the hydrodynamics literature \cite{WaCM2003jcis,FrAT2012sm}. }

Before we supplement the passive model by bioactive terms, we introduce a scaling 
for length, time and energy to write the equations in dimensionless form. \bfuwe{We employ the scales $l=(\tfrac{B}{A})^{1/3}$ and $L=(\tfrac{\gamma}{\kappa})^{1/2} l$ for vertical and horizontal length scales, respectively, where $l \ll L$.} The scales for time and energy are $\tau= L^2 \tfrac{\eta_0}{\kappa l}$ and $\kappa=\tfrac{k_\text{B} T}{a^3}l$, respectively. From now on, we use the dimensionless form of the equations and present all figures in dimensionless scales.
\bfuwe{These can be calibrated to apply to particular biofilm experiments also allowing one to estimate the material constants that have not yet been measured for the complex mixture forming the biofilm. An example of such an estimate is given below in section~\ref{sec:cali}. }

\begin{figure}[h]
\begin{center}
\includegraphics[width=0.65\textwidth]{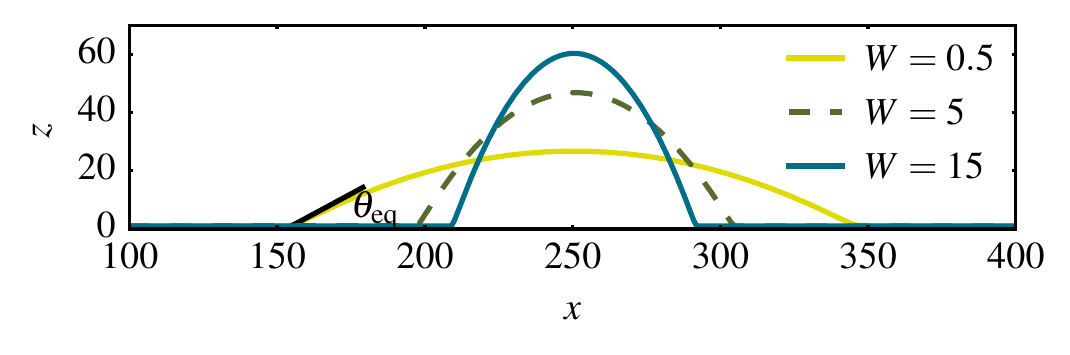}  
\caption{Influence of the dimensionless wettability parameter $W$ [(\ref{eq:ww})] on the shape of equilibrium droplets. Larger values of $W$ result in steeper droplets as $\tan^2 \theta_\text{eq} = \tfrac{3}{5} W$.}
\label{Fig2}
\end{center}
\end{figure}

The resulting dimensionless wettability parameter 
\begin{equation}
W= \frac{A a^3}{l^3 k_\text{B} T}
\label{eq:ww}
\end{equation}
measures the relative strength of the wetting energy \cite{pismen2006asymptotic} as compared to the entropic free energy of mixing. \bfuwe{It is directly connected to the equilibrium contact angle $ \theta_\text{eq}$ of a droplet of a homogeneous mixture 
on the moist agar substrate via the relation 
\begin{equation}
\tan^2 \theta_\text{eq} = \tfrac{3}{5} W \, .
\end{equation}
Larger values of $W$ correspond to a less wettable substrate and result in steeper droplets.} As an illustration, Fig.~\ref{Fig2} shows equilibrium shapes of droplets of a passive mixture on a thin wetting layer of height $h_p=1$ for various values of $W$ as indicated in the legend.

\subsection{Long-wave model for biofilms}
\label{sec:mod-act}

\begin{figure}[htbp]
\begin{center}
\includegraphics[width=0.65\textwidth]{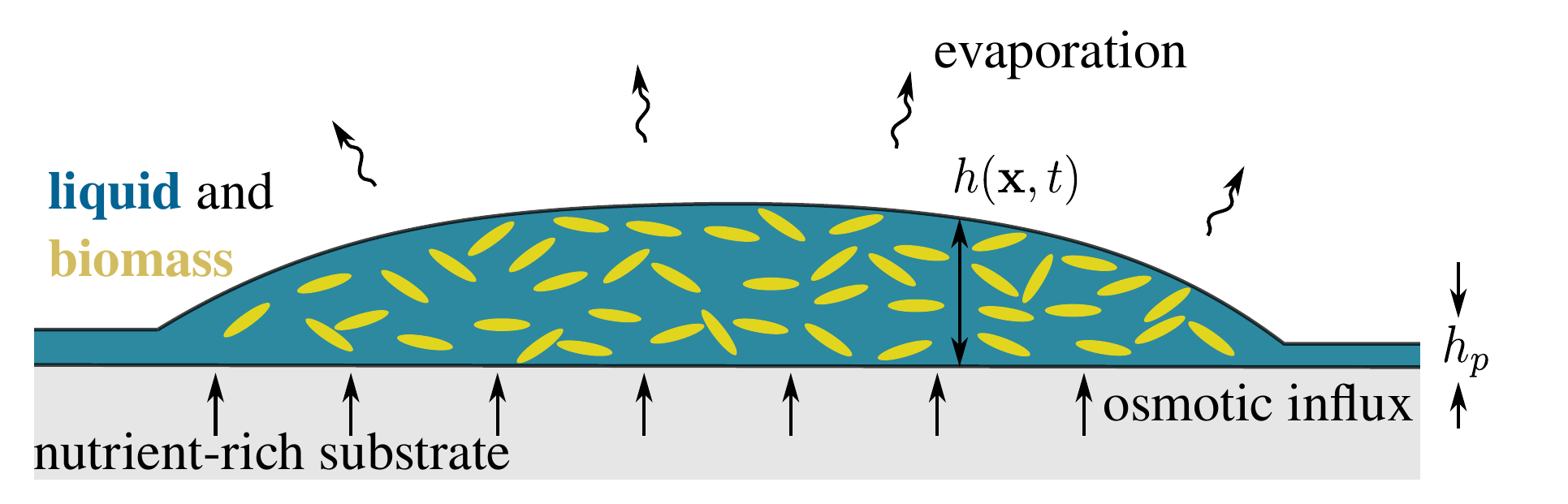}  
\caption{Schematic illustration of the model to study osmotically driven biofilm spreading. A biofilm droplet of height profile $h(\textbf{x},t)$ sits on a thin wetting layer of height $h_p$. Osmotic pressure gradients are generated as bacteria consume water and nutrients to produce biomass via bacterial proliferation and matrix secretion. This causes an osmotic influx of nutrient-rich water from the moist agar substrate into the biofilm. }
\label{Fig3}
\end{center}
\end{figure}

\bfuwe{In the previous section we have presented a gradient dynamics model (\ref{mixtureeqn})-(\ref{vis}) for films of passive mixtures. Now we supplement this model by bioactive processes (proliferation/death) and also account for osmotic influx and evaporation of the solvent. These influences play a crucial role in the spreading dynamics of biofilms \cite{SAW+2012PNASUSA,DTH2014PotRSoLBBS} and are shown schematically in Fig.\ \ref{Fig3}.  From hereon, solvent and solute always represent nutrient-rich water and biomass (bacteria and the extracellular polymeric matrix), respectively. }

\bfuwe{In the course of the studied biofilm spreading mechanism, osmotic pressure gradients are generated as bacteria consume water and nutrient to produce biomass. Then, in response, nutrient-rich water is transported  through the agar-film interface which can not be passed by the biomass. With other words, the osmotic imbalance causes swelling and subsequent spreading of the biofilm as water enters from the moist agar substrate to regulate the fluid concentration. }
Another mechanism that may alter the composition of the biofilm is the evaporation of liquid into the surrounding air. 
As the influx from the moist agar into the biofilm and the evaporation of liquid are mathematically equivalent on our level of description \cite{TTP2010apa}, we lump the two processes into an effective flux term that we call ``osmotic flux'' in the following.
Biomass growth and osmotic flux are incorporated into the model as two non-conserved terms $G(\psi_1, \psi_2)$ and $\zeta(\psi_1, \psi_2) $, respectively. 
This results in evolution equations for the effective layer thicknesses
   \begin{align}
  \partial_t \psi_1 =& \nabla \cdot \left[Q_{11} \nabla \frac{\delta F}{\delta \psi_1} +Q_{12} \nabla \frac{\delta F}{\delta \psi_2}        \right] - G(\psi_1, \psi_2) + \zeta(\psi_1, \psi_2)  \notag \\
 \partial_t \psi_2 =& \nabla \cdot \left[Q_{21} \nabla \frac{\delta F}{\delta \psi_1} +Q_{22} \nabla \frac{\delta F}{\delta \psi_2}        \right] +  G(\psi_1, \psi_2)\, .
 \label{biofilmeqn}
 \end{align}
In the next two subsections we discuss the osmotic flux term $\zeta(\psi_1, \psi_2)$ and the growth term $G(\psi_1, \psi_2)$ in more details. 

\subsubsection{Osmotic influx - properties of $\zeta(\psi_1, \psi_2)$ }
\bfuwe{We assume that the interface between agar and biofilm is semipermeable, i.e., nutrient-rich water can pass through it, but not the biomass nor the agar.
As the bacteria in the biofilm consume water and new biomass is produced, the water concentration is modified from its equilibrium value thereby creating an osmotic imbalance. This imbalance causes an osmotic flux of nutrient-rich water across the interface into the biofilm which tends to equilibrate the osmotic pressure on both sides. When the flat biofilm is in equilibrium with the moist agar substrate for a certain water concentration $\phi_1=\frac{\psi_1}{h}= \phi_{\mathrm{eq}}$,  the osmotic pressure in the biofilm equals that in the agar. }

In general, the osmotic pressure is defined as the variation of the total free energy of a system with respect to its volume at fixed number of osmotically active particles \cite{Doi2011jpcm}. 
Then, in the local description of the present model and taking into account that $\psi_1$ and $\psi_2$ are independent variables, the osmotic pressure in the biofilm is given by the variation of the free energy w.r.t.\ the effective solvent layer thickness at fixed solute layer thickness:
\begin{equation}
  \Pi = \frac{\delta F[\psi_1, \psi_2]}{\delta \psi_1}\Big|_{\psi_2} \, .
\label{eq:piosm} 
\end{equation}
 \bfuwe{Note that for a flat film without wettability influences, this expression reduces to $\Pi = \frac{k_B T}{a^3} \ln\left(\frac{\psi_1}{h}\right)=\frac{k_B T}{a^3} \ln\left(1-\frac{\psi_2}{h}\right)$. In the dilute case of small biomass concentration $\phi_2=\psi_2/h\ll 1$ one obtains the classical linear result $\Pi \approx -\frac{k_B T}{a^3}\phi_2$. However, the general form (\ref{eq:piosm}) is valid for any concentration and mixing energy. }

\bfuwe{The osmotic flux now depends linearly on the osmotic pressure difference between the biofilm and the agar substrate that is considered as a large reservoir with a constant partial solvent pressure $\mu_{agar}$ which guarantees that the influx vanishes for large drops at equilibrium water concentration $\frac{\psi_1}{h} \approx \phi_{\mathrm{eq}}$. This gives the influx
\begin{equation}
\zeta(\psi_1, \psi_2)= - Q_\text{osm} \left(  \frac{\delta F[\psi_1, \psi_2]}{\delta \psi_1} - \mu_\text{agar}  \right ) \, , \label{osm:flux}
 \end{equation}
with $Q_\text{osm}$ being a positive mobility constant. For the particular energy contributions discussed in section~\ref{sec:mod-act} we have 
\begin{equation}
\zeta(\psi_1, \psi_2)= - Q_\text{osm} \left(  \frac{d f}{d\psi_1} -\gamma\Delta (\psi_1+\psi_2) + \frac{k_B T}{a^3} \ln \frac{\psi_1}{h}- \mu_\text{agar}  \right ) \, . \label{osm:flux2}
 \end{equation}
Expressed in $h=\psi_1+\psi_2$ and $\phi_2=\psi_2/h$ this gives
\begin{equation}
\tilde\zeta(h, \phi_2)= - Q_\text{osm} \left(  \frac{d \tilde f}{dh} -\gamma\Delta h + \frac{k_B T}{a^3} \ln (1-\phi_2) - \mu_\text{agar}  \right ) \, . \label{osm:flux3}
 \end{equation}
Recall that in the employed long-wave approximation all gradients perpendicular to the interface are assumed to be small as compared to gradients along the substrate due to the disparity between vertical and horizontal length scales. This implies that concentration variations along the $z$-axis are not captured as such variations in concentrations equilibrate fast. Furthermore, this implies that osmotic fluxes between the substrate and the biofilm on the one hand and evaporation/condensation processes between the gas phase and the biofilm on the other hand are not separated but both contained in Eq.~(\ref{osm:flux3}).
For instance, one clearly sees that the overall influx depends on concentration as well as wettability and capillarity. The dependence of evaporation on surface curvature is sometimes called the Kelvin effect \cite{ReCo2013pre}. }

\subsubsection{Bacterial growth and matrix production - properties of $G(\psi_1, \psi_2)$} 
\begin{figure}[htbp]
\begin{center}
\includegraphics[width=0.65\textwidth]{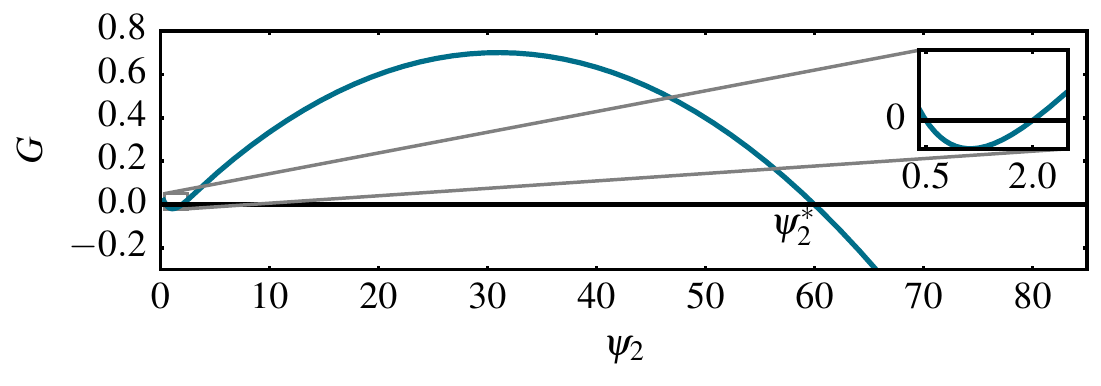}  
\caption{ The bioactivity $G(\psi_1, \psi_2)$ has the form of a modified logistic term and leads to growth if the amount of biomass is smaller then a limiting amount $\psi_2^\star$. Here, $G(\psi_1, \psi_2)$ scaled by the rate constant $g$ is shown for the quasi-steady state of the osmotic influx $\psi_1 = \phi_{\mathrm{eq}} h=\psi_2\phi_{\mathrm{eq}}/(1-\phi_{\mathrm{eq}})$ for $\phi_{\mathrm{eq}}=0.5$.
The inset zooms into the region of small $\psi_2$ and shows the fixpoint structure for small amounts of biomass.}
\label{Fig4}
\end{center}
\end{figure}

\bfuwe{The present section outlines the main argument employed to obtain $G(\psi_1, \psi_2)$ while the details are laid out in the first part of the appendix. }
We assume that the cells grow and secrete the polymeric matrix while consuming the nutrient-rich water. Lumping cell growth and matrix production together into one growth rate constant $g$, 
this process causes a gain term for the biomass $\psi_2$ and a loss term for the liquid $\psi_1$ in the form of a biomolecular reaction proportional to  $g \frac{ \psi_1  \psi_2}{\psi_2+\psi_1}$. The growth of bacteria in a biofilm is not unlimited because processes such as nutrient and oxygen depletion prevent further growth when the biofilm reaches a critical thickness \cite{ZSS+2014NJoP, Dietrich2013}. We consider a functionally simple logistic growth law for the population dynamics that describes a biomass that evolves towards a limiting amount of biomass $\psi_2^\star$ 
 \begin{equation}
 G(\psi_1, \psi_2) = g \tfrac{\psi_1 \psi_2}{\psi_2+\psi_1} (1 - \tfrac{\psi_2}{\psi_2^\star})  \cdot  f_\text{mod}(\psi_1, \psi_2) \, .   
\end{equation}
This growth law is modified locally by $f_\text{mod}(\psi_1, \psi_2)$ for very small amounts of biomass. The term $f_\text{mod}(\psi_1, \psi_2)$ that is further discussed in the appendix introduces a growth threshold at small $\psi_{2}$ corresponding to an unstable fixed point of the growth law and stabilises the wetting layer. The growth threshold is motivated by the fact that at least one bacterial cell is needed for cell division and thus proliferation of biomass does not take place for very small values of $\psi_2$. The resulting bioactive term $ G(\psi_1, \psi_2) $ is shown in Fig.\ \ref{Fig4} assuming a quasi-steady state for the osmotic influx, i.e. the osmotic influx is very fast compared to the biomass growth such that the water concentration in the (flat) biofilm is always at equilibrium with the agar.

\subsection{Numerical treatment}
\label{sec:num}
\bfuwe{
Direct numerical time simulations are performed using a finite element scheme provided by the modular toolbox DUNE-PDELAB \cite{BB2006, BBD+2008C1,BBD+2008C2}. To apply the package, we split Eqs.~(\ref{biofilmeqn}) into four equations of second order in space and write them in a weak formulation for general test functions. The simulation domain for the simulations of 2d (3d) biofilms (i.e., films on 1d (2d) substrates) is $\Omega =  [0, L_x]$  ($\Omega =  [0, L_x] \times [0, L_y]$). It is discretised on an equidistant mesh of $N_x= 512$ ($N_x \times  N_y = 512 \times  512$)  quadratic elements with linear Q1 ansatz and test functions.
On the boundaries, we apply no-flux conditions and set the odd derivatives of $\psi_1$ and $\psi_2$ in space to zero (i.e., in the 1d case $\frac{\partial \psi_{1,2}}{\partial x}=0$ and $\frac{\partial^3 \psi_{1,2}}{\partial x^3}=0$ at $\partial\Omega$).
An implicit second order Runge-Kutta scheme with adaptive time step is used for the time-integration. The resulting linear problem is solved with a biconjugate gradient stabilised method (BiCGStab) and a symmetric successive overrelaxation (SSOR) as preconditioner. The convergence of the solutions has been tested with respect to space discretization. As initial condition we use a small nucleated biofilm droplet at equilibrium water concentration $\phi_{\mathrm{eq}}$. }

\bfuwe{For the analysis of the emerging front solutions, the evolution Eqs.~(\ref{biofilmeqn}) are transformed into the co-moving coordinate system moving with velocity $v$ in which the fronts correspond to steady profiles by adding an advection term to the evolution equations.
Then, continuation techniques \cite{Kuzentsov2013, DWC+2014ccp} are applied, i.e., we directly track the stationary fronts in parameter space. In general, path continuation techniques allow one to obtain steady states of an ordinary differential equation (ODE) by a combination of prediction steps that advance a known solution in parameter space via a tangent predictor and subsequent corrector steps. The latter employ refined Newton procedures to converge to the solution at the next value of the primary continuation parameter. The primary continuation parameter is in our case, e.g., the growth rate constant $g$ in Eq.~(\ref{eq:gg}). As the continuation of the moving fronts is performed within the comoving frame, the front speed itself has also to be 
determined as secondary continuation parameter (or nonlinear eigenvalue) of the system. In this way one may start at an analytically or numerically known solution, continue it in parameter space and obtain a broad range of solutions including their bifurcations and accompanying changes in morphology. For single and coupled thin film equations as Eqs.~(\ref{mixtureeqn}-\ref{vis}) such methods have been successfully employed, e.g., for dewetting two-layer films \cite{PBMT2005jcp}, depinning drops on a rotating cylinder \cite{LRTT2016pf}, dewetting/decomposition patterns in films of mixtures \cite{TTL2013prl} or pattern formation in dip-coating \cite{WTG+2015apa}. For a review see Ref.~\cite{DWC+2014ccp}. The particular software package we employ is \textit{Auto-07p} \cite{DOC+2007, DKK1991ijobac}. Its usage for various thin-film problems is presented in tutorial form in Ref.~\cite{cenosTutorial}. }

If not stated otherwise, throughout the analysis we fix the maximal amount of biomass that can be sustained by the substrate to $\psi_2^\star=60$, the equilibrium water concentration to $\phi_{\mathrm{eq}}=0.5$ and the ratio of the viscosities of biomass and fluid to $\frac{\eta_b}{\eta_0}=20$ and study the biofilm evolution depending on the growth rate constant $g$ and the wettability parameter $W$. The mobility constant of the osmotic influx is first fixed to $Q_\text{osm} =100$ which corresponds to a strong influx that balances the osmotic pressure difference very fast. In section~\ref{front:sol}, we study the influence of $Q_\text{osm}$ on the properties of the biofilm front solutions. \bfuwe{Further sensitivity analyses are found in part two of the appendix.  The relation of the chosen constants, the obtained results and scales and conditions found in experiments are discussed in section~\ref{sec:cali}. } 

\section{Results and Discussion}
\label{sec:res}

In the following, we present and discuss results obtained with the developed biofilm model (\ref{biofilmeqn}). First, we concentrate on the initial growth phase in which the biofilm swells vertically and subsequently spreads horizontally. Then, we focus on the horizontal spreading regime at later times, when the advancing biofilm front is moving with constant shape and velocity.
\bfuwe{Most of our results are obtained for 2d biofilms (i.e., layers on 1d substrates), that correspond to liquid ridges. In the final section \ref{3D} we present also some first numerical results on three-dimensional biofilms (i.e., layers on 2d substrates). }

\subsection{Transition from swelling to spreading}

\begin{figure}[htb]
\begin{center}
\includegraphics[width=0.8\textwidth]{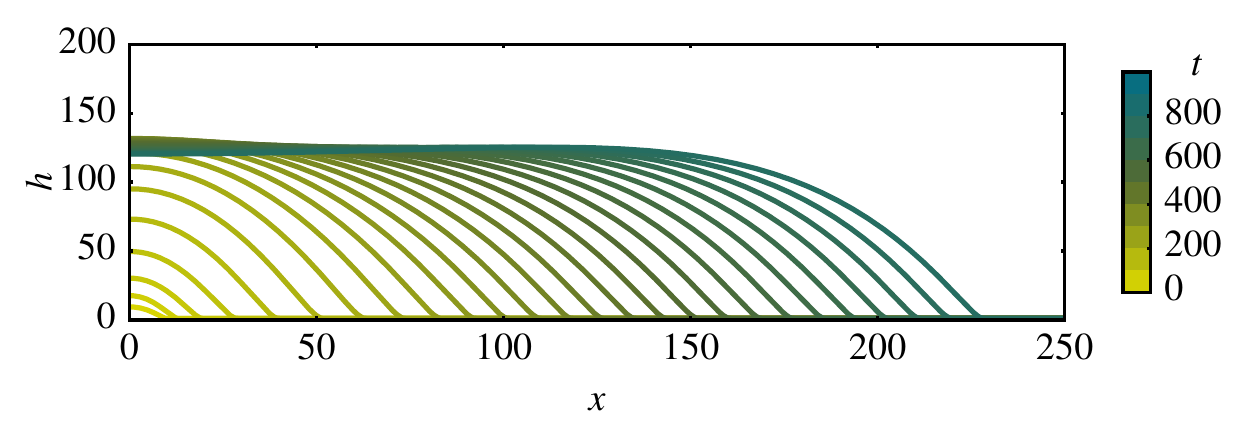}  
\caption{Height profiles $h(\textbf{x},t)$ taken at equidistant times during the early stage of biofilm growth for $g=0.1$ and $W=5$. Only the right half of the symmetric biofilm droplet is shown. The colour of the lines encodes time as indicated in the colour bar.}
\label{FIGBiofilmHeightprofiles}
\end{center}
\end{figure}

In the early stages of growth, the droplet undergoes a transition from initial vertical swelling to subsequent horizontal spreading which is also observed experimentally \cite{SAW+2012PNASUSA,ZSS+2014NJoP}. This behaviour is captured very well by our simple biofilm model (\ref{biofilmeqn}). Fig.\ \ref{FIGBiofilmHeightprofiles} shows height profiles for a growing biofilm at equidistant times. Initially the droplet grows vertically and horizontally. When the biomass approaches the limiting amount $\psi_2^\star$, the vertical growth slows down and only horizontal spreading prevails and provides the film with fresh nutrients and water at the edges.

\begin{figure}[htbp]
\begin{center}
\includegraphics[width=0.9\textwidth]{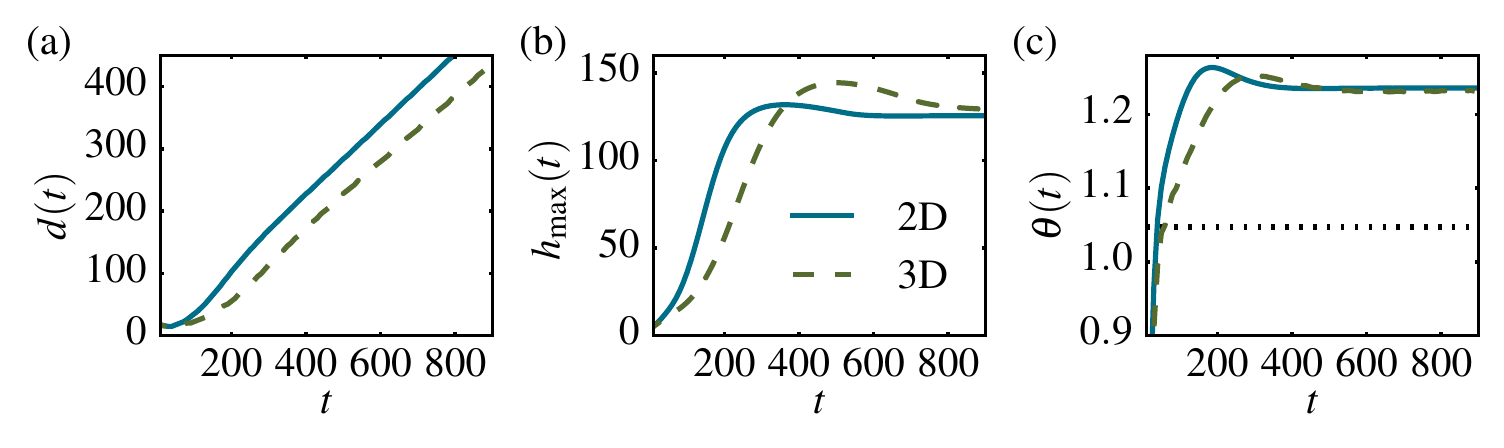} 
\caption{Time evolution of drop size (a), maximal height (b) and contact angle (c) for a two- (solid blue lines) and a three-dimensional biofilm droplet (dashed green lines) with $g=0.1$ and $W=5$. \bfuwe{The horizontal dotted line in (c) indicates the value of the equilibrium contact angle without bioactivity.}}
\label{FIGRadiusSteepness}
\end{center}
\end{figure}

Figure~\ref{FIGRadiusSteepness} shows the transition between vertical and horizontal swelling more quantitatively . 
Shown are the evolutions of the drop size of the biofilm $d(t)$\footnote{We define the size of the biofilm $d(t)$ as the distance between the inflection points of the height profile $h(\textbf{x},t)$at the edges of the droplet.}, the maximal biofilm height and the contact angle $\theta$ \footnote{The contact angle $\theta(t)$ is determined from the slope of $h(\textbf{x},t)$ at the inflection point of the height profile.} with solid blue lines for two-dimensional biofilm growth (in comparison to the three-dimensional case, cf.~section~\ref{3D} below). 
In the initial phase up to $t\approx 80$, the droplet grows slowly, as seen by a slow increase in the drop size $d(t)$ and the maximal film height $h_\text{max}(t)$ [Fig.\ \ref{FIGRadiusSteepness} (a) and (b)]. However, in this phase vertical growth is more dominant than horizontal spreading since the contact angle $\theta(t)$ [Fig.\ \ref{FIGRadiusSteepness} (c)] is still increasing. After the initial phase the size of the colony is nearly growing linearly with time. Both, the maximal 
film height and the contact angle $\theta(t)$ are increasing steeply, \bfuwe{reach a maximum before they decrease again slightly to approach their plateau values at time $t\approx 400$. Beyond this time the biofilm is growing by a growth front that advances horizontally with a constant velocity and a constant front profile, i.e., the dynamic contact angle is also constant. Note that the plateau value for the contact angle correspond to stationary values that are larger than the equilibrium contact angle as expected for an advancing front [see dotted horizontal line in Fig.~\ref{FIGRadiusSteepness}(c)]. }

The above described sequence of events is in accordance with the experimental finding \cite{SAW+2012PNASUSA,ZSS+2014NJoP}. But in contrast to earlier theoretical models for osmotic spreading of biofilms \cite{WCM+2011,WK2012JEM, SAW+2012PNASUSA, DTH2014PotRSoLBBS} the wetting properties of the biofilm on the agar substrate are explicitly included, allowing us to describe the horizontal spreading of the biofilm on a wetting layer.

\subsection{Front solutions}
\label{front:sol}
\begin{figure}[h]
\begin{center}
\includegraphics[width=0.9\textwidth]{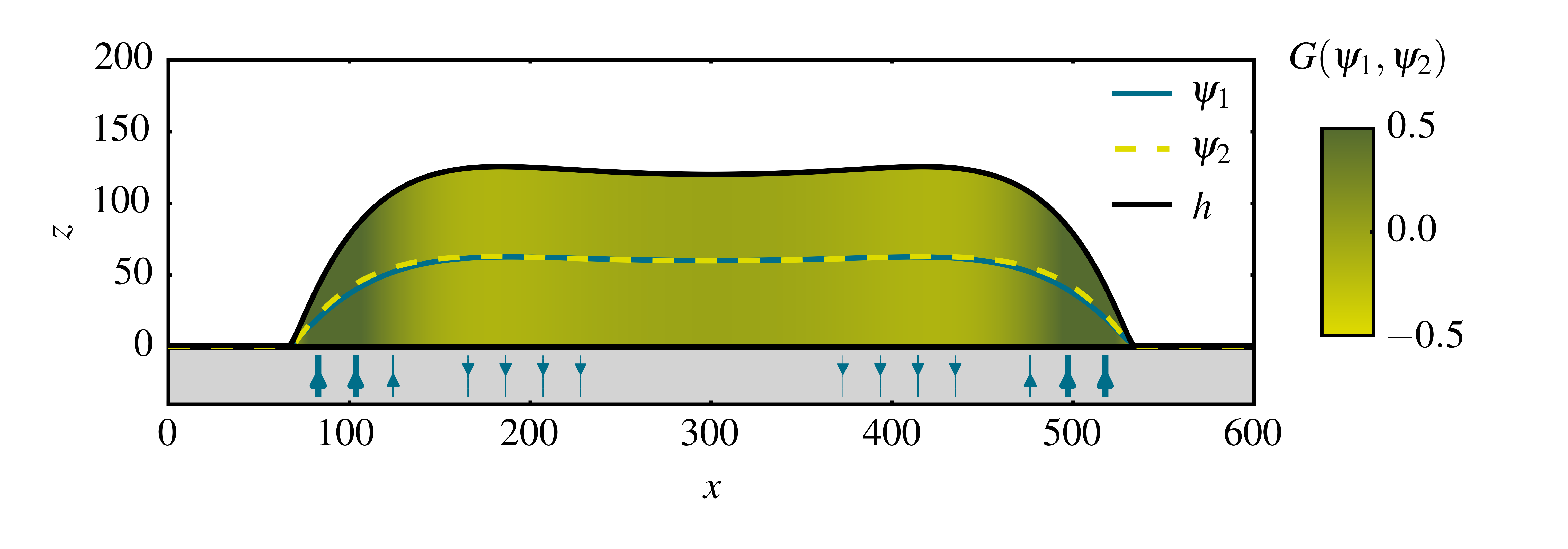}  
\caption{Bioactivity and osmotic influx for an exemplary biofilm front profile obtained as a snapshopt at $t=800$ in a simulation for $g=0.1$ and $W=5$. The colouring of the droplet indicates the bioactivity $G(\psi_1, \psi_2)$. The biomass growth is strongest at the edges of the film. The direction and thickness of blue arrows below the film represent the direction and strength of the effective osmotic flux term $\zeta(\psi_1, \psi_2)$. The influx is strongly positive at the edges of the biofilm, where biomass growth causes an osmotic imbalance decreasing  the solvent concentration below its equilibrium value, i.e.\ $\frac{\psi_1}{h}<\phi_{\mathrm{eq}}$.}
\label{FIGFrontDetailed}
\end{center}
\end{figure}

\bfuwe{Fig.\ \ref{FIGFrontDetailed} shows an example of a growing biofilm at a late time, after the transition in the dynamics from swelling to horizontal spreading, when the biofilm front advances with constant velocity and shape. }
At its center, the biofilm is in a steady state: the biomass has reached the limiting value $\psi_2^\star$ and the water concentration is $\phi_{\mathrm{eq}}$, so that \bfuwe{biomass production and degradation are at a dynamic equilibrium as are osmotic influx and evaporation of water, i.e., no net growth or influx take place ($G=0$ and $\zeta=0$).} The effective layer thicknesses of liquid $\psi_1$ and biomass $\psi_2$ that are represented in Fig.\ \ref{FIGFrontDetailed} by the solid and dashed lines, respectively, are constant in the central region of the biofilm. At the edges, the biomass has not yet reached the limiting amount. The biomass growth, which is in Fig.\ \ref{FIGFrontDetailed} encoded in the colouring of the biofilm, alters the water concentration from the equilibrium value $\phi_\text{eq}$ causing an osmotic influx of water from the agar substrate into the biofilm. The direction and strength of the influx are in Fig.\ \ref{FIGFrontDetailed} indicated by the blue arrows underneath the biofilm.
 
\begin{figure}[htb]
\begin{center}
\includegraphics[width=0.48\textwidth]{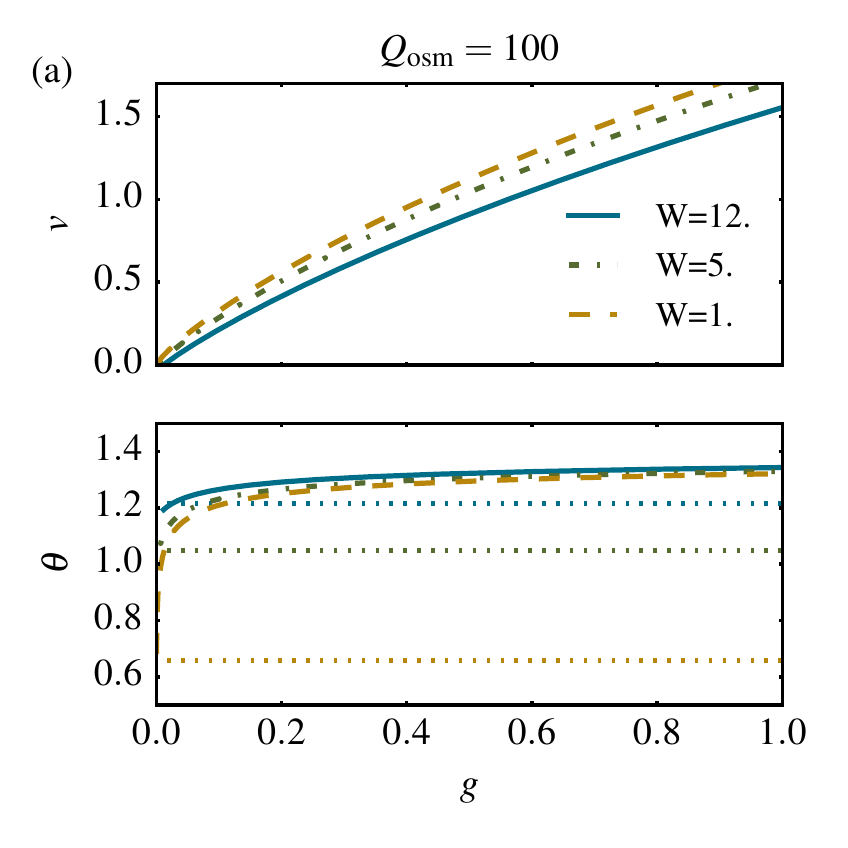}  \includegraphics[width=0.48\textwidth]{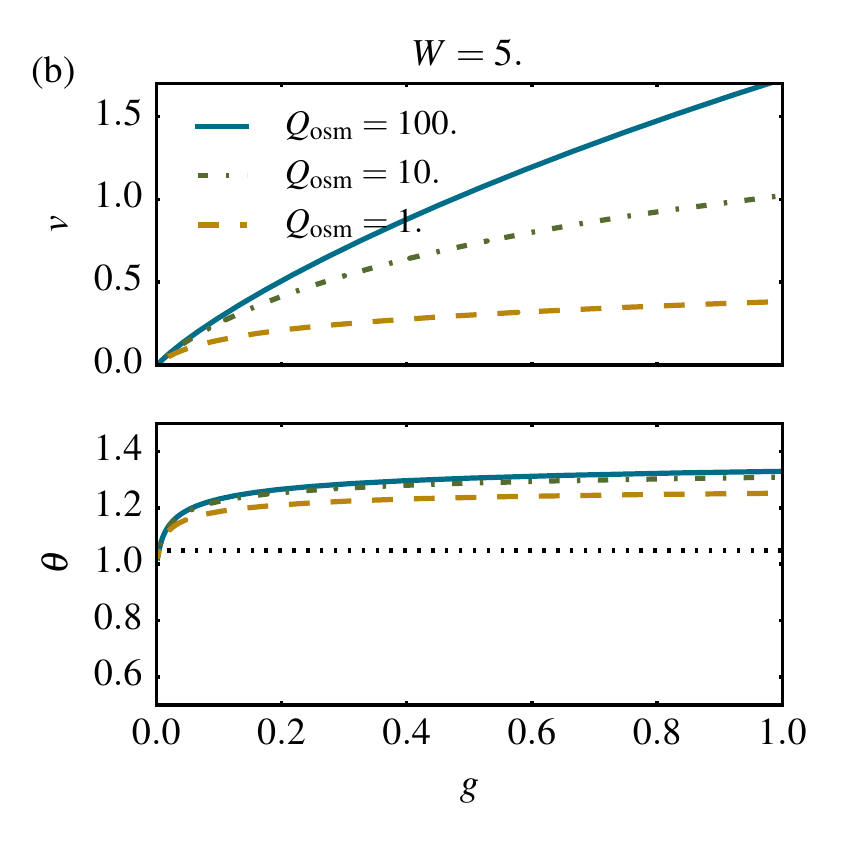}  
\caption{Front velocity (upper row) and contact angle (lower row) depending on the bioactivity rate constant $g$ obtained by parameter continuation. (a) shows the influence of the wettability parameter $W$ at fixed $Q_\text{osm} =100$. In (b), the value of $Q_\text{osm} $ which determines the strength of the osmotic influx is varied at fixed $W=5$. The dotted horizontal lines indicate the equilibrium contact angles of the corresponding passive mixtures obtained with Eq.~(\ref{eq:ww}). }
\label{FIGFronts}
\end{center}
\end{figure}

We use continuation techniques to further analyse the front solutions (see section~\ref{sec:num}). Fig.\ \ref{FIGFronts} shows the dependencies of the front velocity $v$ and dynamic contact angle $\theta$ (determined from the slope at the inflection point of the profile) of the stationary advancing fronts on the biomass growth rate constant $g$. As expected, the front velocity (Fig.\ \ref{FIGFronts}, top row), increases with $g$ because the biomass growth causes the osmotic imbalance in the biofilm and triggers the subsequent influx of water. 

The dynamic contact angle of the advancing biofilm front also increases with increasing $g$ (Fig.\ \ref{FIGFronts}, bottom row).  If the front advances very slowly for small bioactivity rates $g$, the contact angle approximately corresponds to the equilibrium contact angle of the passive mixture (dotted lines) given by Eq.~(\ref{eq:ww}). If a stronger biomass growth pushes the biofilm front forward, the front profile is steeper. \bfuwe{This indicates that the dynamic contact angle increases with front velocity as known from the passive advance of fronts or drops, e.g., for spreading drops or sliding drops on an incline \cite{Bonn2009,TVNB2001pre}.}  The spreading is slower for larger values of $W$ which correspond to steeper passive droplets, i.e., the agar substrate is less wettable for the biofilm (Fig.\ \ref{FIGFronts}, left column), albeit this effect is small. The front also slows down if the osmotic influx is weaker which can be modeled by smaller values of $Q_\text{osm}$ corresponding to a smaller 
permeability of the interface between moist substrate and biofilm (Fig.~\ref{FIGFronts}, right column).  Recall that the model is analysed in the dimensionless long-wave form and $l \ll L$ holds for the dimensional vertical and horizontal length scale. Consequently, the dimensional front profiles are much more shallow than in the dimensionless representation of the profiles shown in our figures.

\subsection{Three-dimensional biofilm growth} \label{3D} 
So far, we have analysed the behaviour of two-dimensional biofilms, which correspond actually to biofilm ridges. While the extension of the model to 3d biofilms does not represent any conceptual difficulty, the use of, for example, continuation techniques is much more technically involved.
The swelling and spreading dynamics of \bfuwe{radially-symmetric}  3d biofilm droplets qualitatively resembles the 2d case as can be seen from the evolution of the drop size $d(t)$ (i.e., here the radius), the maximal biofilm height $h_\mathrm{max}(t)$ and the contact angle $\theta(t)$ in Fig.~\ref{FIGRadiusSteepness} (dashed green lines). The transition from a swelling droplet towards a centrally flat film with an advancing front is slower than for the 2d biofilm but the front velocity and the constant dynamic contact angle of the evolving front profile are approximately equal to the values for the 2d case. Note that in both cases the front velocity approaches a constant value.

\begin{figure}[hbt] 
\begin{center} 
\includegraphics[width=\textwidth]{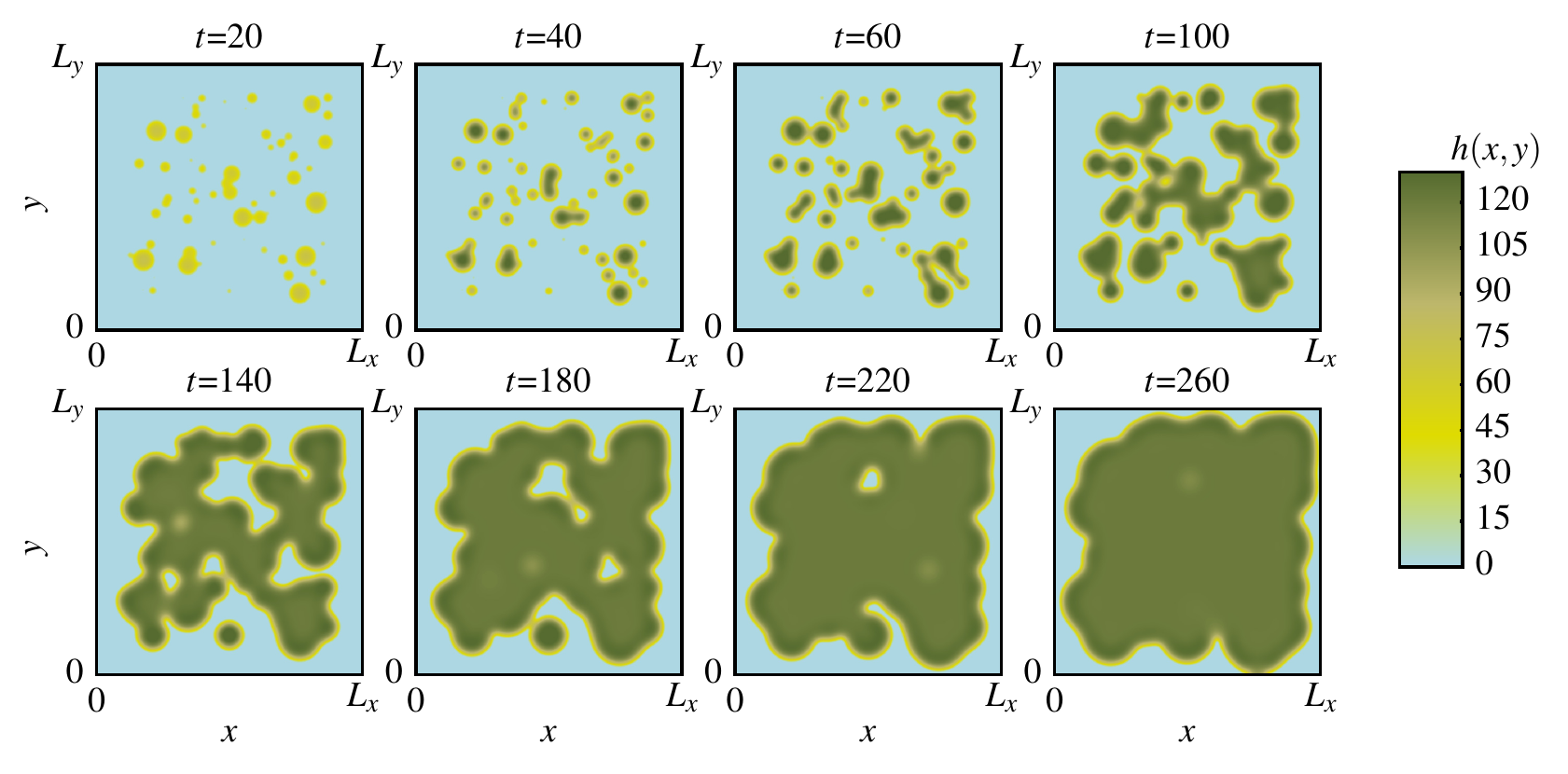} 
\caption{Simulation of a growing three-dimensional biofilm (i.e., on a two-dimensional substrate) with $g=0.3$, $W=1.5$, $ \frac{\eta_b}{\eta_0}=10$ and $L_x=L_y=2000$. Starting from small nucleated biofilm droplets, a homogeneous biofilm evolves. } 
\label{Fig9} 
\end{center} 
\end{figure}

As an illustration for 3d biofilm growth we show in the following an example result obtained from the direct numerical simulation of Eqs.\ (\ref{biofilmeqn}).  Starting from randomly distributed small droplets, the time simulation presented in Figure \ref{Fig9} shows an evolution towards an extended biofilm. The individual droplets first swell vertically and subsequently spread horizontally as observed in the 2d case. If droplets touch, they merge and form larger aggregates until eventually, an extended biofilm at equilibrium water content emerges.  

\subsection{Experimental calibration} 
\label{sec:cali} 

\bfuwe{What we have presented so far is a qualitative analysis of osmotic biofilm spreading employing the evolution equations introduced in section~\ref{sec:mm}. We are well aware that our model is quite simple and neglects many aspects of biofilm complexity. Furthermore, some parameters as, e.g., the mobility coefficients for the transfer of water from the agar into the biofilm, are not known. However, in the following we discuss how to calibrate the relevant scales and dimensionless parameters using the experiments by Seminara and coworkers \cite{SAW+2012PNASUSA}. As the biofilm evolution is highly sensitive to the environmental conditions, e.g., the limiting biofilm height varies between several tens to several hundred $\mu$m \cite{SAW+2012PNASUSA,ZSS+2014NJoP}, the model has to be calibrated separately for each individual experiment. Here we specifically use the experiment in Ref.~\cite{SAW+2012PNASUSA}.}

\bfuwe{From the biofilm height of $\sim$30\,$\mu$m, the typical time scale for the transition between vertical and horizontal swelling of about 10\,h and the typical spreading velocity of about 200\,$\mu$m $h^{-1}$ we estimate the vertical length scale $l$=250\,nm, the horizontal length scale $L$=20\,$\mu$m and the time scale $\tau$=6\,min. From the measured dynamic contact angle of about 1$^\circ$, together with the estimated length scales we obtain the dimensionless wettability parameter $W=5$, which is the value we have used for most calculations. A typical dimensionless growth rate constant $g=0.1$ (Figs.~5, 6, and 7) corresponds to a dimensional growth rate of 1\,h$^{-1}$, a reasonable assumption for bacterial cell division. }

\bfuwe{The estimate for the transfer coefficient of water between the agar substrate and the biofilm is not an easy task, since it crucially depends on the permeability of the agar and the interface thickness. In our simulations we have assumed that osmotic fluxes are faster than or on the same time scale as biomass growth and therefore the time scale of spreading is dominated by the interplay of biomass growth and surface forces. Using the relation for the dimensionless mobility $Q_\mathrm{osm}=\chi k_BT\tau/(\mu l L a^3)$, where $\chi=3000$\,nm$^2$ \cite{DTH2014PotRSoLBBS} denotes the permeability of the agar, $\mu=10^{-3}$\,Pa\,s denotes the dynamic viscosity of water and with the size of the solute $a=0.1\dots1\,\mu$m we find a range of values for $Q_\mathrm{osm}=1\dots1000$, i.e., about the range we have studied in Fig.~\ref{FIGFronts}. }

\section{Conclusion}
\label{sec:conc}

\bfuwe{We have developed a simple model for the osmotic spreading of growing biofilms on the surface of moist agar substrates under air. To do so we have first laid out that a film of passive (not bioactive) mixtures or suspensions can be described by two coupled thin-film equations for the effective layer heights of solvent and solute. These can either be written in the form of a gradient dynamics on an underlying energy functional as derived in the literature or in the more usual hydrodynamic form of coupled thin film and advection-diffusion equations. The former way is advantageous as it allows for an easy extension towards more involved interaction and interfacial energies that it incorporates automatically in all relevant transport channels including the osmotic influx and evaporation. }

\bfuwe{We have then shown that the passive model can serve as a basis for the study of the growth dynamics of biofilms as it can be supplemented by additional terms that account for bioactive processes. Namely, these are proliferation/death processes, production of intercellular polymeric matrix and also the passive osmotic influxes caused by the growth of biomass in the film. We have derived and discussed a particular, simple model to study the osmotically driven spreading of biofilms which is observed experimentally for certain bacterial strains at moist agar-air interfaces. The model focuses on the fluid-like character of young biofilms and especially incorporates the wetting behaviour and the fluid flow in a consistent way by introducing a wetting energy (i.e., a Derjaguin or disjoining pressure) into the dynamic model. To our knowledge this has not been done before. }

\bfuwe{
  Within this framework, the biofilm spreads on a stable wetting layer, whereby spreading is driven by biomass production and a resulting osmotic influx from the agar. This has offered us the opportunity to study the influence of passive material parameters, such as, the wettability of the agar substrate and surface tension of the biofilm-air interface on the spreading behaviour of the biofilm. 
For the early stage of growth, the model reproduces the transition from initial vertical swelling to horizontal spreading. 
In the long-time limit, the model shows that biofilms grow laterally by stable stationary growth fronts. These have been further analysed by applying continuation techniques. In this way, we were able to characterize the influence of model parameters, such as, the growth rate constant of the bacteria, the wetting properties and the strength of the osmotic influx on the velocity and profile of the fronts.}

\bfuwe{Our results show that our approach results in a qualitatively correct and quantitatively reasonable simple asymptotic model of osmotic biofilm spreading. It should be seen as a parallel approach to complex multi-scale biofilm models that can normally only be tackled by expensive numerical simulations. Therefore, we plan to incorporate in the future further features of biofilm growth. Possible extensions of the model include, for example, the auto-production and dynamics of surfactants in the biofilm which would allow one to study the role of surface tension gradient-driven flows (Marangoni flows) on the spreading behaviour \cite{ARK+2009PNASU}. }

\section*{Appendix}
\subsection*{Model for bacterial growth and matrix production}
\bfuwe{
We assume that biomass is produced in a bimolecular reaction by consuming the nutrient-rich water. The production of biomass includes the growth of bacteria as well as the production of extracellular matrix. The time evolution of the concentrations of water $\phi_1 = \frac{\psi_1}{h}$ and biomass $\phi_2 = \frac{\psi_2}{h}$ due to this processes is then given by
\begin{align}
& \partial_t \phi_1 = - g \phi_1 \phi_2 \\
& \partial_t \phi_2 = \alpha g \phi_1 \phi_2 
\end{align}
where $\alpha$ is a factor that describes the efficiency of this transformation which describes for example the amount of nutrients that the bacteria need to sustain their metabolism. For simplicity, we set $\alpha=1$. The growth rate constant $g$ depends, e.g., on the nutrient concentration in the agar substrate and the bacterial strain. 
Translating this growth into an evolution for the effective heights of water and biomass yields
\begin{align}
& \partial_t  \psi_1= - g \frac{\psi_1 \psi_2}{\psi_1 + \psi_2} \\
& \partial_t \psi_2 = g \frac{\psi_1 \psi_2}{\psi_1 + \psi_2} \, .
\end{align}
The growth of bacteria and the production of extracellular matrix is not unlimited because it is limited by effects as the availability of nutrient and oxygen across the biofilm. These processes limit the biofilm to a maximal height  that corresponds, e.g., to a balance of nutrient diffusion and consumption \cite{ZSS+2014NJoP,Dietrich2013}. As it is beyond the scope of this work to model oxygen and nutrient concentration directly, we introduce a limiting amount of biomass $\psi_2^\star$ that can maximally be sustained by the substrate,  which depends, e.g., on the nutrient concentration in the agar. This can be achieved by introducing a factor $(1 - \tfrac{\psi_2}{\psi_2^*} )$ into  the growth term, corresponding to the well known logistic law, that describes a population dynamics which evolves towards a limiting value \cite{Murray1993}. }

\bfuwe{In addition, we introduce a growth threshold at a small value $\psi_{2,u}=2.0$ that is motivated by the fact that at least one bacterial cell is needed for cell division and matrix production. It is introduced such that it shifts the onset of the logistic growth to this value. 
\begin{align}
& \partial_t  \psi_1= - g \frac{\psi_1 (\psi_2 - \psi_{2,u})}{\psi_1 + \psi_2} \left(1 - \frac{\psi_2}{\psi_2^\star} \right) \\
& \partial_t \psi_2 = g \frac{\psi_1 (\psi_2 - \psi_{2,u})}{\psi_1 + \psi_2} \left(1 - \frac{\psi_2}{\psi_2^\star} \right)  \, .
\end{align}
Recall that the biofilm coexist with a thin wetting layer of height $h_p$ on the moist agar substrate. Growth of bacteria in the wetting layer away from the biofilm would correspond to the spontaneous creation of new biofilm droplets away from the original biofilm. Here we avoid this by modifying the growth term. We introduce a stable fixed point at $\psi_{2,s}=\phi_{eq} h_p =0.5$ in the growth term, which ensures that no bacterial growth takes place in the wetting layer that shall be in osmotic equilibrium with the agar. The final growth term is then given by
 \begin{align}
 G(\psi_1, \psi_2)    &= g \frac{\psi_1 (\psi_2 - \psi_{2,u})}{\psi_1 + \psi_2}   \left(1 - \frac{\psi_2}{\psi_2^\star} \right) \left( 1 - \exp(\psi_{2,s}- \psi_2)   \right) \\
 &= g \frac{\psi_1 \psi_2}{\psi_2+\psi_1} (1 - \frac{\psi_2}{\psi_2^\star}) \underbrace{(1 - \frac{\psi_{2,u}}{\psi_2})  \left( 1 - \exp(\psi_{2,s}- \psi_2)   \right)}_{f_\mathrm{mod}(\psi_1,\psi_2)}   \, .   
\end{align}
Please note that the introduction of the growth threshold and the stabilisation of the wetting layer only ammends the growth term at very small amounts of biomass as with increasing $\psi_2$, the term $f_\mathrm{mod}(\psi_1,\psi_2)$ fast approaches one. Outside the zoom shown in the inset of Fig.~\ref{Fig4} the logistic growth law remains practically the same. Other choices of $f_\mathrm{mod}(\psi_1,\psi_2)$  with the same fixed point structure do not change the qualitative behaviour of the system. }

\subsection*{Sensitivity analysis for $\psi_2^*$ and $\phi_\mathrm{eq}$}

\begin{figure}[htbp]
\begin{center}
\includegraphics[width=0.49\textwidth]{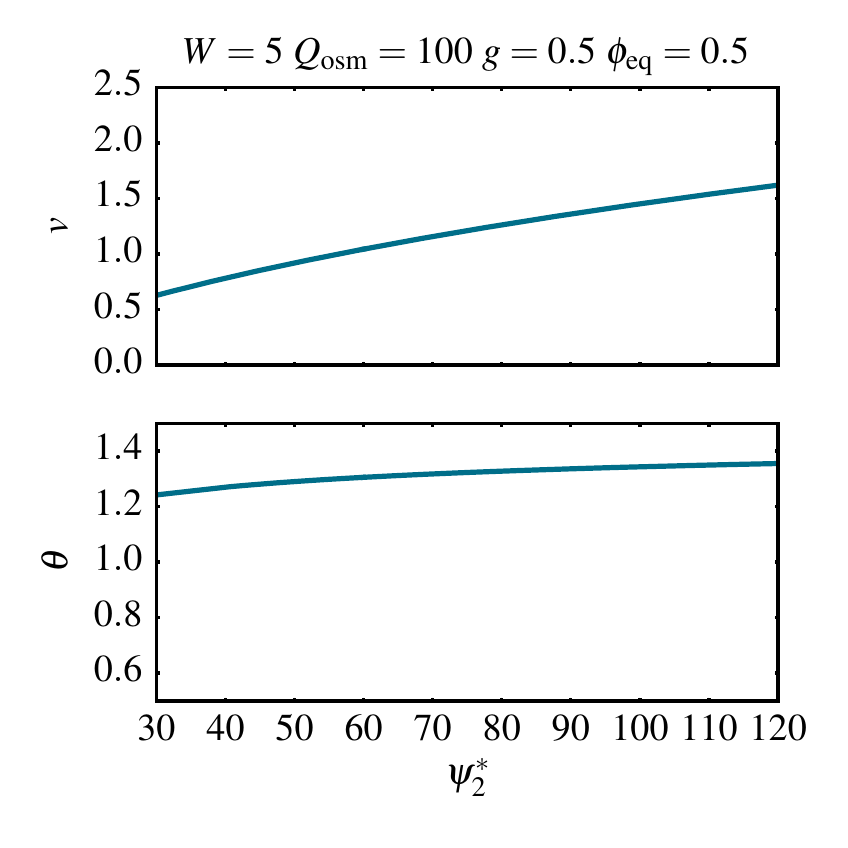}  
\includegraphics[width=0.49\textwidth]{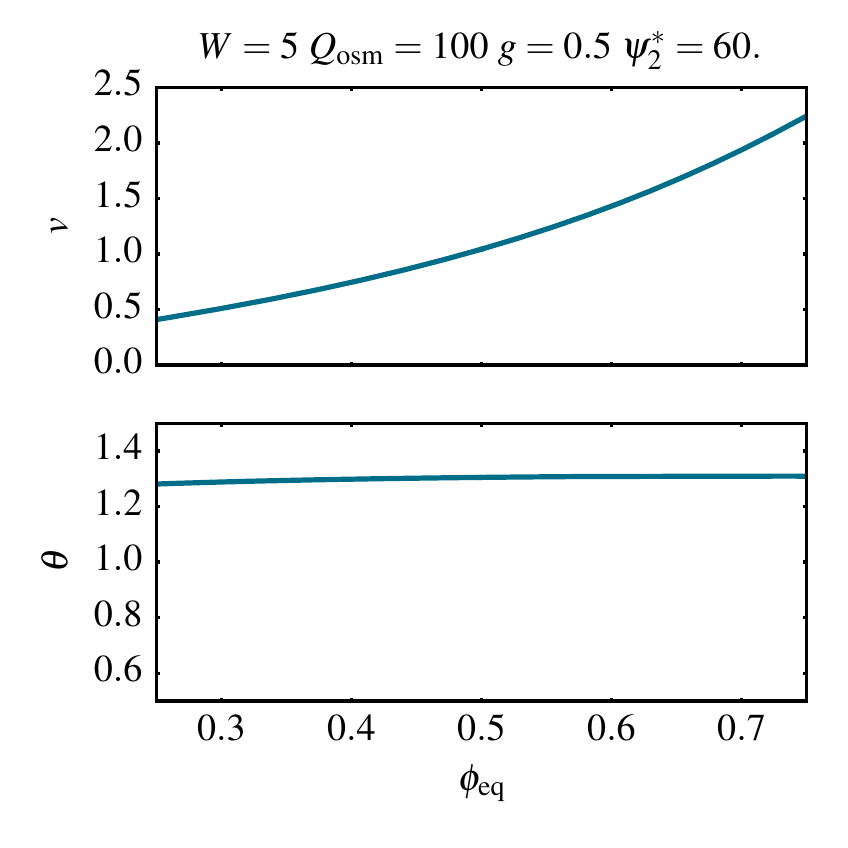} 
\caption{Velocity and contact angle of the biofilm fronts depending on the limiting amount of biomass $\psi_2^*$ and  the equilibrium water concentration $\phi_\mathrm{eq}$.}
\label{FigSUP}
\end{center}
\end{figure}

To study the sensitivity of the model on the less confident parameters as the limiting amount of biomass $\psi_2^*$ and  the equilibrium water concentration $\phi_\mathrm{eq}$, these parameters are varied over a broad range of values. The qualitative behaviour of the model does not change but the velocity of the biofilm speading depends on these parameters as shown in Figure \ref{FigSUP}.

\section*{Acknowledgments}
We are grateful for the financial support by the German Academic Exchange Service (DAAD) and Campus France (PHC PROCOPE grant N$^\circ$ 35488SJ). The LIPhy is part of the LabEx Tec 21 (Investissements de l'Avenir, Grant Agreement No. ANR-11-
LABX-0030). We thank Sigol\`ene Lecuyer of the LIPhy, CNRS / Universit\'e Grenoble-Alpes and the anonymous reviewers for fruitful discussions and critical input, respectively.

\section*{Conflict of Interest}
All authors declare to have no conflicts of interest in this paper.

\end{document}